\documentstyle[emulate_apj]{article}
\slugcomment{{\it ApJ}, in press}
\lefthead{S. Andreon}
\righthead{Deep $K_s$ luminosity function at z=0.3}
\begin{document}
 
\title{Deep near--infrared luminosity function of 
a cluster of galaxies at z=0.3\footnote{Based on observations
collected at the European Southern Observatory, Chile, ESO N° 62.O-0369
and, in part, on observations with the NASA/ESA Hubble Space Telescope}
}
\author{S. Andreon}
\affil{Osservatorio Astronomico di
Capodimonte, Via Moiariello 16, 80131 Napoli, Italy\\
E-mail: andreon@na.astro.it}
 
\begin{abstract} 
The deep near--infrared luminosity function of AC\,118, a cluster of galaxies at 
$z=0.3$, is presented. AC\,118 is a bimodal cluster, as evidenced both by our 
near--infrared images of lensed galaxies, by public X--ray {\it Rosat} images 
and by the spatial distribution of bright galaxies. Taking advantage of the 
extension and depth of our data, which sample an almost unexplored region in the 
depth vs. observed area diagram, we derive the luminosity function (LF), down to 
the dwarf regime ($M^*+5$), computed in several cluster portions. The overall 
LF, computed on a 2.66 Mpc$^2$ areas ($H_0=50$ km s$^{-1}$ Mpc$^{-1}$), has an 
intermediate slope ($\alpha=-1.2$). However, the LF parameters depend on the 
surveyed cluster region: the central concentration has 
$2.6^{+5.1}_{-1.7}$ times more bright galaxies and 
$5.3^{+7.2}_{-2.3}$ times less dwarfs per typical galaxy
than the outer region, which includes galaxies at an average projected 
distance of $\sim580$ kpc (errors are quoted at the 99.9 \% confidence level). 
The LF in the secondary 
AC\,118 clump is intermediate between the central and outer one. In other words, 
the near--infrared AC\,118 LF steepens going from high to low density regions. 
At an average clustercentric distance of $\sim580$ kpc, the AC\,118 LF is 
statistically indistinguishable from the LF of field galaxies at similar 
redshift, thus suggesting that the hostile cluster environment plays a minor 
role in shaping the LF at large clustercentric distances, while it strongly 
affects the LF at higher galaxy density.
\end{abstract}
 

\newpage
 
\section{Introduction}

The luminosity function (LF hereafter), i.e. the number density of galaxies 
having a given luminosity, is critical to many observational and theoretical
problems (see e.g. Binggeli, Sandage \& Tammann 1988). For example, it is needed 
to interpret galaxy counts (e.g. Driver \& Phillipps 1996), to compute the 
spatial covariance from the angular correlation function (e.g. Koo \& Kron 
1992), and to constrain the primordial power spectrum of density fluctuations 
(e.g. Ostriker 1993). From an observational point of view, the LF is the natural 
``weight" of several quantities since most statistical quantities are weighted on the 
relative number of objects in each luminosity bin. For example, the fraction of 
blue galaxies, which is the basic measure used for measuring the Butcher--Oemler 
effect (Butcher \& Oemler 1978, 1984), is a (normalized) convolution, over the LF, 
of the color distribution at a given luminosity. Furthermore, due to the role 
played by luminosity in the inclusion of the objects in the studied sample 
(faint objects are often excluded or under--represented), the knowledge of the 
LF is fundamental to the calculation of the selection function and it is needed 
to derive the actual galaxy properties from the measured quantities. For 
instance, the large increase of the brightness of galaxies with $z$ may be partially 
due to a $z$--dependent sampling of the LF (Simard et al. 1999).

Due to the central role played in many astrophysical problems, the optical LF 
of galaxies in clusters has been extensively studied (e.g., to cite just a few 
papers dealing with large numbers of clusters, Gaidos 1997; Valotto et al. 1997; 
Lumsden et al. 1997; Garilli, Maccagni \& Andreon, 1999). However, not all 
wavebands carry the same information: optical filters are, with respect to 
near--infrared ones, more affected by short lived starburst events and thus better tracers 
of the metal production rate but worse tracers of the underlying stellar mass 
(Bruzual \& Charlot 1993). It is therefore valuable to extend the LF measure to 
other bands. Near--infrared filters are useful in several aspects: with respect 
to optical filters they are less affected by internal and Galactic absorption, 
and their differential (type to type) K--corrections are quite small (up to 
$z\sim1$) thus not altering the cluster content just because of an observational 
effect. 

The present knowledge of the cluster near--infrared LF is quite fragmentary and 
it is limited either to small areas at moderate depth (e.g. Barger et al. 1996; 
Trentham \& Mobasher 1998; Aragon-Salamanca et al 1993) or to relatively large 
areas, but at bright luminosities (e.g. de Propris et al. 1999; Barger et al. 
1998). This implies, for instance, that dwarfs galaxies are not sampled outside 
the cluster core. As a first step to overcome these drawbacks, we present in 
this paper the LF of a cluster of galaxies, AC\,118, computed over one of the 
largest areas (in Mpc$^2$) and to a depth only rarely achieved for any cluster 
of galaxies.

AC\,118, also known as Abell 2744 (Abell 1958; $z=0.308$), is one of the most 
observed clusters at intermediate redshift. It was first studied by Couch \& 
Newell (1984) on photographic plates, then low dispersion spectroscopic data 
were acquired (Couch \& Sharples 1987; Barger et al. 1996). For the 
very central region of the cluster ($r<\sim90"$) {\it Hubble Space Telescope} 
images have been used for morphological studies (Couch et al. 1998; Barger et 
al. 1998) and mass determination through gravitational lensing experiments 
(Smail et al. 1997). Near--infrared ($K'$-band) photometry (Barger et al. 1996) 
of the very central region of this cluster at intermediate depth and with a 
coarse angular resolution ($FHWM\sim1.7$ arcsec) is also available. Barger et 
al. (1996) present $U$ and $I$ photometry for the brightest $K'$ sources. 
AC\,118 shows an excess of blue galaxies (Couch \& Newell 1984), commonly known 
as Butcher--Oemler effect (Butcher \& Oemler 1978, 1984) and it is one of the 
clusters that triggered the discussion on whether the star formation rate 
increases or decreases during the galaxy infall in the cluster (Barger et al. 
1996; Balogh et al. 1997, 1998).

The paper is organized as follows: in the next section, the data and the data 
reduction are presented. In Section~3, the AC\,118 LF is derived in a few 
cluster regions. The discussion and a summary are presented in Section~4.

In this paper we assume $H_0=50$ km s$^{-1}$ Mpc$^{-1}$ and $q_0=0.5$.
If $\Omega_\Lambda=0.7$, $\Omega_{M}=0.3$, 
AC118 galaxies would be $\sim0.3$ brighter. The main result of this paper,
a clustercentric dependence of the LF, is independent on the cosmological parameters
because all compared galaxies are at the same distance from us.

\section{Data and data reduction}

\subsection{AC\,118 Observations}

AC\,118 observations were carried out at the 3.5 m NTT with SOFI (Moorwood, Cuby 
and Lidman, 1998) on October, 11, 1998, in the frame of an observational program 
aimed at deriving the Fundamental Plane at $z\sim0.3$. SOFI is equipped with a 
$1024\times1024$ pixel Rockwell ``Hawaii" array. In its large field mode 
the pixel size is 0.292 arcsec and the field of view $5\times5$ arcmin. 
The field was observed in the near--infrared $K_s$ passband ($\lambda_c=2.2 
\mu$; $\Delta\lambda \sim 0.3 \mu$) during a photometric night with
good seeing ($FWHM<0.8$ arcsec). The total useful exposure time is 15900 s, 
resulting from 
the coaddition of many short jittered exposures. Jittering was controlled by the 
automatic jitter template (described in the SOFI User Manual) which produces a 
set of dithered frames. Offsets were randomly generated within a box of 
$40\times40$ arcsec centered on the cluster center ($\alpha: \ 00 \ 14 \ 19, 
\delta: \ -30 \ 23 \ 18$, J2000). The effective exposure time of each individual 
AC\,118 frame was 1 min in $K_s$, resulting from the average of 6 exposures of 
10 s each. Photometric calibration has been obtained by observing a few standard 
stars, interspersed with AC\,118 observations, taken from the 
list of Infrared NICMOS Standard Stars now published in Persson et al. (1998). 
The effective exposure time for standard stars was 100 s, given by exposures 
taken at five different array locations, each one being the average of 10 
integrations 2 s long. Fig. 1 shows the final $K_s$ image of AC\,118.

\subsubsection{Reset, gain and illumination corrections}

All images have been flat--fielded by flaton--flatoff. All pixels whose gain 
differs from the average by more than 30 \% have been flagged and not 
used in the image combining.

Hawaii chips have a special feature: after reset, the zero--level is not 
constant over the field and its amplitude depends on the total charge collected 
in the previous exposure. The variation of the zero--level is due to the fact 
that rows have not the same reset voltage after reset because resetting an 
Hawaii chip is a power demanding operation (Mackay et al. 1998). This pattern is 
additive and equal for all pixels in the same row since they are reset in 
parallel. This pattern is unimportant for science images, whose background is 
anyway spatially not uniform and variable in time. However, this spatial pattern 
and its time variation are important in the determination of the gain correction 
(i.e. of the flaton--flatoff). Fig 2 shows two typical zero-level patterns, and 
also their difference. This difference is $\sim150$ ADU with variations of 
$\sim50$ ADU from one row to another, which induces, if not accounted for, a 
photometric zero--point variation on the field at a few percent level. The 
zero--level has been measured using appropriate images ({\it special} 
flaton--flatoff frames), as suggested by C. Lidman (private communication, now fully described 
in the SOFI manual). In the determination of the zero--level it is assumed that 
during the observations the lamp is stable; this assumption has been verified a 
posteriori, by comparing of a few series of such measurements.

In accurately reduced images
the flux of a standard star should not depend on its location.
In order to test the accuracy of the flat--fielding, a standard 
star has been observed 16 times, displacing the telescope pointing between each 
exposure by $\sim80$ arcsec, in such a way that the standard star appears on a 
grid $4\times4$, and its flux at the different positions on the 
chip has been measured. Two more stars were in the field of view and were also
used for this test. 
Figure 3 shows the deviation from an ideal response (i.e. 1 everywhere), 
after gain correction and before any other correction. The measured 
flux is well within 1 \% from 1 in most of the sampled locations. The RMS 
deviation is 0.7 \% which, if not corrected for, would induce a 0.007 mag error. Part 
of the observed scatter is due to photometric errors on individual magnitude 
measurements, accounting for 0.005 mag. In just one location the measured 
flux differs by 2.4 \%. However, this outlier value could be partly due to a 
transient slightly hot/cold pixel un--recognized as such and un--corrected for. 
We recall that transient hot/cold pixels are recognized and corrected in the 
next phase, which combines images taken at different pointings.

Since the RMS deviation from the mean is $\sim0.7$\%, our images do not require a 
supplementary illumination correction, which is instead often considered in the 
reduction of 
near--infrared images. As a comparison, the RMS deviation from the mean of the 
NICMOS 1 \& 2 on {\it Hubble Space Telescope} is $\sim2$\% (Colina, Holfeltz \& 
Richie, 1998) and no illumination correction is included in the NICMOS pipeline 
reduction, to our best knowledge.

\subsubsection{Background removal}

For the background subtraction a user--friendly software tool, Eclipse 
(Devillard, 1997), has been used.

In the near-infrared, observers are faced with changes of the intensity, spectrum 
and spatial shape of the sky. Nodding the telescope modulates the 
astronomical objects intensity (on a give pixel) more quickly than the background 
variation.

From a technical point of view, a high--pass filter will remove the slower 
changing background and will leave untouched the higher frequence astronomical 
signal. To make this operation effective, Eclipse scales time--adjacent images to
the mean (or median) of the image that must be background--subtracted. In doing
this operation, Eclipse assumes that the variation of the background is 
multiplicative and coherent over the whole field during the considered short time
interval. Then, images are low--pass filtered in the time--line direction, using a
filter 11 exposures large centered on the image whose background is to be
subtracted, allowing a rejection of the faintest and of the three brightest pixels in
order to take into account the existence of celestial objects, cosmic rays and
intermitting hot/cold pixels. With this choice of the Eclipse settings, more bright 
pixels than faint ones are clipped during the sky determination. This is
the natural choice when, as in our case, sky pixels contaminated by sources have
an asymmetric intensity distribution, i.e. when the number of sources with positive
counts exceeds the number cold pixels. While pixel
masking would be preferable, it is not available within the Eclipse software.

This operation gives the background which is subtracted to the original flat- 
fielded image. We want to stress that no multiplicative scaling is applied to the image 
to be background--subtracted. We verified, by means of Midas pipelines developed 
by the author for the analysis of near--infrared images of the Coma cluster 
(Andreon et al. 2000), that Eclipse correctly performs this complex task.

\subsubsection{Photometric calibration}

The airmass coefficient is computed by means of the science frames, due to the 
fact that AC\,118 has been observed at several hour angles. We determine the 
airmass coefficient from the measure, on 190 out of 265 frames, of the apparent 
flux of a small galaxy (a possible early--type galaxy of AC\,118) in an 
uncrowded region. The upper panel of Figure 4 shows the airmass dependence of the 
instrumental magnitude of the considered galaxy. The adopted value for the 
atmospheric absorption, 0.08 mag airmass$^{-1}$, is compatible with that derived 
from the observations of the (admittedly few, 5) photometric standard stars 
observed during the night, and is equal to the value assumed by Persson et al. 
(1998) in defining standard magnitudes for the standard stars used here. Notice 
that the residual scatter in the data, after correction for airmass differential 
extinction, is 0.03 mag for the considered galaxy, of which 0.03 mag (i.e. 
almost all) are due to photometric errors in individual magnitude measures. The 
similarity of the expected and observed scatters gives a direct confirmation that 
the observing night was photometric, and that the data reduction is accurate.

Aperture magnitudes have been used to measure the apparent flux of the 
standards, using the same aperture adopted by Persson et al. (1998) for the 
photometric standard stars, i.e. 10 arcsec in diameter. The zero—point scatter 
derived from standard stars is 0.008 mag, of which 0.005 mag are due to 
photometric errors associated to the Persson et al. (1998) photometry.


\subsubsection{Combining the images and final details}
 
Images must be coaligned before coadding. The jitter data reduction 
procedure uses cross--correlation techniques in the image domain to measure 
offsets with subpixel precision, adopting, as first estimates, the telescope 
spatial offsets written in the header of the FITS files. Images have been 
combined using the task {\tt imcombine} under IRAF, since Eclipse does not fit 
our needs. Files containing the flux scaling (since AC\,118 has been observed 
through different airmasses), the bad pixel mask and the relative spatial 
offsets were given in input. The task gives in output the composite image, the 
image of the measured dispersion among the input images and the number of pixels 
used (i.e. the exposure map in units of 1 minute). As ``mean" for the {\tt 
imcombine} task, we adopt a straight average allowing the rejection of up to one 
high and one low pixel values, to allow for the presence of cosmic rays or 
intermittent hot/cold pixels. We remember that {\tt imcombine} deals only with 
integer shifts, and for this reason spatial offsets are rounded to the nearest 
integer. This fact inflates the final PSF by $\sim0.5$ pixels.
 
Finally, a bright saturated star produces on the final image a row brighter than 
average (and a symmetric ghost row in the opposite image quadrants). A constant, 
computed as described in Andreon (1993) has been subtracted to these two rows.

The sky brightness ranged from 12.9 to 13.2 mag arcsec$^{-2}$ during the night. 
The seeing, in the combined image, is 0.75 arcsec (FWHM, $\simeq2.5$ pixels). 
The sky noise in the fully exposed part of the image is 24.0 mag arcsec$^{-2}$.


\subsection{Control field: Hubble Deep Field South}

Public images for the Hubble Deep Field South (HDF-S) have been taken in $K_s$ 
with SOFI at NTT (Da Costa et al. 2000), i.e. with the same instrument, filter 
and telescope as the AC\,118 image. This ensures an almost perfect homogeneity 
of the data. A few characteristics of HDF-S images are listed in Table 1, 
together with those relative to AC\,118 observations. Public images are already 
reduced. The reduction of these data follows the same lines described for 
AC\,118 and makes use in large part of the same software used by us, i.e. Eclipse. The 
claimed photometric calibration has an error of 0.1 mag for HDF-S-1 and 0.05 mag 
for HDF-S-2 (Da Costa et al. 2000). 
 
Differently from AC\,118 images, HDF-S images have been spatially resampled and 
filtered during the combining phase, and for this reason the noise in the final 
image looks smaller than it really is. As a consequence, magnitude errors in Da 
Costa et al. (2000) are underestimated. In order to perform an uncorrelated 
measurement of the noise, as we did for AC\,118 images, we binned the HDF-S images, 
thus reducing the correlation between adjacent pixels. Table 1 presents the 
fully corrected noise in AC\,118, HDF-S-1 and HDF-S-2. Once corrected for 
correlated noise, the noises in the HDF-S and AC\,118 images are consistent with the 
exposure times. Galaxy counts computed from these images agree with literature 
ones (Da Costa et al. 2000), as confirmed also by Figure 5.


\subsection{Detection and completeness magnitude}

Objects has been detected by using SEx version 2.1 (Bertin \& Arnouts 1996).
For AC\,118 we made use of the RMS map for a clean detection. Due to the
varying exposure time across the field of each image, due to the dithering, we
consider here only the central square areas listed in Table 1. 

Galaxies have no well defined edges, therefore their luminosity depends on how galaxy
edges are defined. We adopt Kron magnitudes (see Kron 1980 for the exact
definition, and Bertin \& Arnouts 1996 for a software implementation), defined 
as the flux integrated in an area with size adapted to each galaxy. 
Unfortunately, Kron magnitudes depend sensibly on the determination of the object
size, which is very difficult for faint objects and in crowded regions, such as
the core of AC\,118. Therefore, aperture magnitudes are adopted for faint objects.
In detail, as a measure of the magnitude of the galaxies,  magnitudes computed
within 2.5 Kron radii are adopted for galaxies brighter than $K_s=18$ mag and
aperture magnitudes (at the 4.4 arcsec aperture, which correspond to 24 Kpc) are
used for fainter galaxies. Figure 5 shows that the two quantities give almost
identical galaxy counts in a large magnitude  range, and in particular near the
bridge magnitude ($K_s=18$ mag). Since the LF is computed by using galaxy counts,
and not individual object magnitude, it is insensitive to differences between Kron and aperture
magnitudes of each individual object.

In computing galaxy counts it is a standard practice to correct them for missing 
flux, i.e. for the flux not within the isophote or aperture. Usually, this
correction is applied to all galaxies, independently on their luminosity. There is
no reason for applying this correction for the LF determination, and furthermore,
this correction has two shortcomings in our case. First of all, the absolute
magnitude distribution of the galaxies of the same apparent magnitude differ in
the two compared directions, because the sample in the cluster line of sight
is the superposition of a
volume complete (the cluster) and a flux limited sample (the fore--background)
while the control field is a flux limited sample. Thus, the luminosity
correction is not the same, even at a fixed apparent magnitude, due to both
cosmological effects (cosmological dimming and K correction at least)
and the surface brightness profile of galaxies which depends on
the (unknown) absolute luminosity. 
Second, even
under the optimistic hypothesis that corrections are perfectly known
for each individual object, the catalog completeness is undermined by {\it bright} 
galaxies with low surface brightness
(a long and thorough discussion on this topic is presented in Andreon \&
Cuillandres, 2000).

The magnitude completeness is defined as the magnitude where objects start to
be lost because their central brightness is lower than the detection threshold.
It is measured as the brightest magnitude of the detected galaxies having
central brightness equal to the detection threshold (see Garilli, Maccagni \&
Andreon 1999 for details).  For AC\,118, the ($5\sigma$) limiting magnitude is
$K_s\sim20.5$ mag in a $4.4$ arcsec aperture. Of course, many 
fainter and smaller objects are  visible on the image, because the limiting
magnitude is fainter using a smaller  aperture, such as the ``standard" $3$
arcsec aperture, or a $3$ FWHM aperture.

\subsection{Star/galaxy classification}

Most previous similar studies have near--infrared images whose resolution is too 
coarse for allowing object classification as star or galaxy from the extent of
the sources. Thus, observers were obliged to adopt the classification performed
either on optical images of the cluster (e.g. Barger et al. 1996) or 
colors (e.g. De Propris et al. 1999). Given the good seeing and sampling of the
AC\,118 $K_s$ image, the star/galaxy classification can be based on the object
extent, as measured in the near--IR image itself, by adopting the SEx
star/galaxy classifier. The very central area of AC\,118 has been observed by
the {\it Hubble Space Telescope} (Couch et al. 1998). Only a few stars (as
classified by their compactness in the {\it HST} image) are brighter than the
$K_s$ completeness limits (AC\,118 is at high Galactic latitude, see Table 1),
and all of them are correctly classified as stars using the $K_s$ image. Only a
few galaxies, out of hundreds, are misclassified as stars because of their
compactness. Therefore, {\it HST} confirms the goodness of our ground--based
star/galaxy classification.

\subsection{Comparison to literature data}

AC\,118 has been observed in the $K'$ band by Barger et al. (1996). Figure 6
shows the comparison between their $K'$ magnitudes, computed in a 5 arcsec
aperture, and our $K_s$ mag, computed in a 4.4 arcsec aperture for common
objects. The line shows the ``one--to--one" relation, and it is not a fit to
the data. The agreement is good. Inspection of our image shows that outliers
to the right of the ``one--to--one" relation are pairs of objects, blended
in Barger et al. (1996) and resolved here as separate objects, due to
better seeing and sampling. A few of these objects are 
shown as inset in Figure 1.

\section{The AC\,118 luminosity function}

The final image of AC\,118 shows features which likely are gravitational arcs
(Y. Mellier, private communication), some of which are 
previously unknown. The brightest two are magnified
in the low panel of Figure 1. Other suspected lensed galaxies, not visible on 
the {\it HST} image because they fall outside its field of view, are likely present, but 
their confirmation requires a full lensing analysis of the image, which is 
outside our aims. The arcs visible in the near--infrared image
and the other visible on the {\it HST} image
point out two main mass concentrations: a central one, 
and another one offset to the North--West, thus confirming the binary structure 
suggested by the spatial distribution of the galaxies (Figure 1). Inspection of 
public {\it Rosat} HRI \& PSPC images shows a similar binary structure, whose 
brighter clump is coincident with the assumed center of AC\,118, while the second 
clump is at the North--West.

Figure 7 shows galaxy counts toward HDF-S and AC\,118. Magnitudes have not been 
corrected for Galactic extinction, because the correction is very small ($<\sim 
0.01$ mag and in any case negligible with respect to the HDF-S photometric 
zero point error). Counts in the cluster direction are larger than in the field 
direction and the difference is large.

The cluster LF is the statistical difference between counts in the cluster 
direction and in the control field direction (Zwicky 1957, Oemler 1974). 
Its error should account for 
Poissonian fluctuations of counts along the cluster and field lines of sight and 
for non--Poissonian fluctuations of the counts, due to the non--zero galaxy 
correlation function. 
We take into account the last term according to Huang et al. (1997). 
Thus, the statistical significance of any claim on the LF does not assume that
fore/background in the cluster line of sight is the ``average" one (or
that observed in the control field), but instead takes into full account that
the background counts fluctuate, from region to region, and fluctuate more than
$\sqrt n $. We stress 
that background fluctuations enter twice in the error budget because the LF is 
given by a difference of two galaxies counts, each one subject to background 
fluctuations. For simplicity, we assume that errors of different nature can be 
added in quadrature (i.e. that standard error propagation laws can be used), and 
in the error propagation we neglect the difference between Poissonian and 
Gaussian distributions.

In the very center of the cluster, galaxy density is high. Thus, in order to 
perform an accurate measure of the LF in this area, a crowding correction must 
be computed. For each magnitude bin, we consider as the area usable for the 
detection the one which is not filled by brighter galaxies. 
Since galaxies have no well defined edges, we 
assign to each galaxy an excluded area equal to half their isophotal 
($\mu\sim24$ mag arcsec$^{-2}$) area. This correction turn out to be 
significant only in 
the central region of AC\,118. Results are robust with respect to the 
implemented crowding correction.

Figure 8 shows the AC\,118 global LF, i.e. those computed using all the $\sim 5
\times 5$ arcmin field of view, corresponding to 2.66 Mpc$^2$ at the AC\,118
redshift. The best fitting Schechter (1976) function is also plotted. In order to
take into account the amplitude of the bin, we fit the data with a Schechter
function convolved with the bin width. The best fit  parameters are: $K^*_s=15.3$
mag ($M_{K^*_s}=-26.0$ mag) and $\alpha=-1.2$, where  $\alpha$ is the slope of the
faint part of the LF, and $M^*$ is the knee of the LF, i.e. the magnitude at
which the LF starts to  decrease exponentially. The confidence levels of the best
fit Schechter  parameters are plotted in Figure 9. The AC\,118 global LF is
smooth, well  described by a Schechter function ($\chi_{\nu}^2=0.7$) and has an
intermediate  slope, down to $K_s=20.5$ mag ($M_{K_s}=-20.8$ mag), which is
$\sim5$ mag below  $M^*$, well in the dwarf regime \footnote{For the time being,
we use the term ``dwarf" for galaxies  whose magnitude is $M>\sim M^*+4$}.

Figure 10 summarizes two relevant features of the published cluster LFs in the 
near--infrared: it shows the magnitude limit reached as 
a function of the area surveyed. Points in the lower--right corner of the diagram 
are the most informative about the LF, because of the large area coverage and of 
the depth reaching the dwarf luminosity, but they are also the most expensive in 
terms of telescope time. The present study is marked by the 
point nearest to the lower--right corner of the graph. With respect to previous 
determinations, we explore at the same time one of the largest area and one of 
the deepest LFs, mainly because SOFI has 16 times more pixels, on average, than 
many previous instruments and because observations have been tailored for 
Fundamental Plane studies, which require to measure the surface brightness 
profile of galaxies, not just their integrated magnitude. Thus, present data 
offer the possibility to study the environmental dependence of the LF and to 
sample possible effects on dwarf galaxies, in contrast to previous near-infrared cluster surveys that explored
either smaller areas or a smaller magnitude range.

In order to search for a dependence of the LF on the environment, we compute the LF 
in two regions, centered on the two clumps of AC\,118 (see Figure 1). Each 
region has an area of $\sim 0.5$ Mpc$^2$, i.e. the typical area observed at 
comparable depth (see Figure 10). The two LFs are plotted in Figure 11, together 
with their best Schechter fits. A Kolmogorov--Smirnov test shows that 
{\it data points} differ at 99.5 \% confidence level (i.e. they differ at 
$\sim3\sigma$, but it should be remembered that this test makes no use of errors). 
The main clump of AC\,118 is poorer in dwarfs and is richer in 
bright galaxies than the North--West clump. Figure 12 shows the 
confidence levels of the best fitting Schechter parameters. {\it 
Assuming that the data are drawn from a Schechter function}, they differ at 
almost $95$ \% confidence level (i.e. $\sim2\sigma$, but now errors are taken 
into account). This plot confirms the 
previous finding: $\alpha$ is shallower in the main clump than in the NW clump, 
indicative of an environment poor in dwarfs. 

We now compare the LF computed in the main clump of AC\,118 with that in the 
outer region. The outer region excludes both the two $\sim 0.5$ Mpc$^2$
regions  centered on the two clumps and also a no--man--land around the NW
clump (see Figure 1). The  galaxies in the main clump have an average
projected distance of
180 kpc, while  the galaxies in the outer region are 580 kpc away, on average,
and 1.1 Mpc at  most. Thus, the outer region is still well inside the cluster,
as confirmed also by the fact that the galaxy overdensity is large enough for
the LF to be computed with the differential counts method. Figure 13 shows that
the inner and outer LFs  differ (at more than 99.99999 \% confidence level
according to a  Kolmogorov--Smirnov test): there are far more dwarfs and
somewhat fewer bright galaxies per typical galaxy (say $K_s\sim18$ mag) in the
outer region. In order to quantify these excesses, galaxies having $K<17,
17<K<19, K>19$ mag are defined bright, typical and dwarf galaxies,
respectively. These break magnitudes correspond, at the cluster redshift, to
$M_K=-24.2$ and $M_K=-22.2$ mag, respectively. We found that the number of
dwarfs per typical galaxy in the outer region exceeds the one observed in the
central region by a factor $5.3^{+7.2}_{-2.3}$, where errors are quoted at
99.9\% confidence level and are computed according to Gehrels (1996), i.e. by
taking into account that the error on this ratio is partly binomial and partly
Poissonian. The excess of bright galaxies in the main clump, with respect to
the NW region, is instead $2.6^{+5.1}_{-1.7}$  (99.9\% confidence level).  

Figure 14 gives the confidence levels of the best fitting Schechter parameters
for the inner and outer region. They are located in quite different parts of
the diagram and point out the same differences shown in Fig. 13. The same
figure also shows that $K^*$ is undetermined in the outer region. In other
terms, the outer region LF can be accurately described by a powerlaw.

There are some caveats in comparing values of one single  best fit parameter,
say $M^*$, when they are computed by means of a fit with a function with three
free parameters: a superficial inspection of the plotted confidence  contours
or of Table 2 would lead to a fairly different conclusion: $M^*$ is brighter in
the outer region and therefore bright galaxies are more abundant there, in
apparent contradiction with our previous claims. However, errors on best fit
parameters are strongly coupled and one should be cautious when interpreting
changes in one parameter, say $M^*$, when the two other parameters ($\alpha$
and $\phi^*$) change at the same time. Since parameters are coupled, we
emphasize the direct comparison of the data itself.

The excess of dwarfs in the outer region is large enough to make the number of dwarfs
per typical galaxy in this region larger than the one observed for the global
LF (which includes the outer region): we found $2.8^{+3.7}_{-1.5}$ more dwarfs here
than over the whole area. With respect the SW region, the figure is $2.6^{+3.9}_{-1.5}$.
 
In conclusion, the near--infrared luminosity function of AC\,118 depends on the 
considered cluster location: in the inner central region, dense in galaxies, 
there are more bright galaxies and less dwarfs per unit typical galaxy, than in 
the NW clump or in the outer region. In other words, the AC\,118 LF steepens 
going from high to low density regions. The outer region is the richest, among 
the three considered, in dwarfs and, at the same time, the poorest in very 
bright galaxies. The found differences among LFs measured in different cluster
regions cannot be due to variations of 
background counts among the various cluster lines of sight because we have
fully included in the error budget this source of error (which is the largest 
one).

\subsection{Comparison to literature LF}

Barger et al. (1996) present the $K'$ (virtually indistinguishable from $K_s$ 
for our purposes, see Figure 6) composite LF of the very central region of three 
intermediate redshift clusters at $z\sim0.3$, one of which is AC\,118. Their LF 
reaches $K'=19$ mag, i.e. 1.5 mag brighter than our magnitude limit (compare B96 and 
our points in Figure 10). They have in their composite sample $\sim300$ member 
galaxies to be compared to the $\sim500$ cluster galaxies of in our sample of 
AC\,118 alone. 
Their best fit parameters are $K'=15.74\pm0.13$ mag and $\alpha=-1.0\pm0.12$. 
Since Barger et al. (1996) observed only the central region of the cluster, we 
consider here only the central area of AC\,118 and we compare their best fitting 
values to those we derived for the central clump (which of course includes just a 
fraction of our total sample of galaxies). We computed confidence levels for the 
best fit of the composite LF of Barger et al. (1996) reading their data and 
errors from their Figure 7. Figure 15 presents our confidence levels for the 
main clump of AC\,118 and for the composite sample of Barger et al. (1996). The 
two $1\sigma$ confidence levels largely overlap, thus meaning that the two LFs 
are equal within the errors. This is expected, due to the good 
agreement between individual magnitudes of galaxies in common shown in Fig. 6 
\footnote{Barger et al. (1996) quote errors of the best fit parameters without 
specifying for how many free parameters they are. The comparison of our Fig. 15 
to their errors shows that their quoted errors are for one free parameter. 
Adopting the usual convention of quoting error for the number of interesting 
parameter of the fit (see e.g. Avni 1976 or Press et al. 1993), which in the 
present case is two, i.e. $\alpha$ and $M^*$, the actual error on the slope of the 
LF is about twice larger than claimed by Barger et al. (1996).}.

The $K_s$--band tightly maps the rest--frame $H$--band emission for objects at 
$z\sim0.3$. Thus, without making any assumption on the K--correction value, we can 
compare the $H$ band zero-redshift LF of the Coma cluster (De Propris et al. 
1998, Andreon \& Pell\'o 2000) to our $K_s$ band LF of AC\,118. 

The needed transformation is:

$M_{H_{z=0}}=-(m-M)+K_{z=0.3}+0.09$

The two Coma LFs are measured on different parts of the Coma cluster: De Propris
et al. (1998) studied a $\sim1$ Mpc$^2$ area around the cluster center and found a
shallow slope ($\alpha=-1.0$), with an hint of steepening at faint  magnitudes.   
Andreon \& Pell\'o (2000) instead studied a $\sim0.6$ Mpc$^2$ area
offcentered by $\sim0.4$ Mpc and found an overall slope of
$\alpha=-1.3$.  Both works found $M_H^*\sim-24.6$ mag. For AC\,118, we found
$K_s^*=15.3$ to  $16.8$ mag depending on the considered area, which corresponds to
$M_H^*=-26.2$  to $-24.7$ mag, in good agreement with Coma $M_H^*$ when errors are
taken into  account. The trend for a LF shallower in the central region than in
the  other more external region suggested for Coma by Andreon \& Pell\'o (2000) is
confirmed for AC\,118, with the major difference that data and analysis are
homogeneous for the latter cluster, whereas for the  former are not.

Figure 16 compares the LF computed in the outer region of AC\,118 to the field
LF, measured at $0.2<z<0.6$ (Cowie et al. 1997). These LFs are measured in two 
very different ways: the cluster LF is computed on a volume--limited sample, 
whereas the field LF is computed on a flux--limited sample (in fact, the Cowie 
et al.' sample in not actually flux--limited, because it is incomplete and has a 
complex selection function). The two LFs, both concerning galaxies at 
intermediate redshift, have very similar shapes, and both conspicuously differ  
from the LF of the inner region of AC\,118 shown in Figure 13. The similarity of 
these two LFs corroborates both the change of the LF parameters with the
environment and the measured shape of field LF, which is based on a sample with 
a complex selection function. Environment seems to have played no role in modifying the 
near--infrared LF in the outer region of AC\,118, because of the similarity of 
the LFs in the field and in the AC\,118 outer region. Instead, Figure 13 shows 
that the environment modifies the LF shape at larger galaxies density or smaller 
clustercentric distances.

\section{Discussion}

The main result of this study is that the LF shape depends on location within the
cluster. 
Among the studied regions, the cluster center, i.e. the main clump ($d\sim180$ 
kpc), has many more bright galaxies per unit typical galaxy (say 
$K_s\sim18$ mag, $M_H\sim-23.5$ mag), as shown in the previous section. The 
converse holds for dwarfs, which are less numerous in the center than elsewhere. 
The other observed clump, the second one in X--ray luminosity, presents LF 
characteristics intermediate between the central clump and the outer region. At 
an average clustercentric distance of $\sim580$ kpc, the AC\,118 LF is statistically
indistinguishable, whithin the present errors, from the field LF at similar redshift. 

Our definition of the main clump is sound: X-ray images and gravitational 
lensing distorsion of background galaxies point out two main concentrations, of 
which the brightest in X--ray is also the one which we call main. Thus, our definition of 
main and center are independent and therefore unbiased by the presence of bright 
galaxies.

There is not anything like 
a near--infrared LF of AC\,118, since it depends on the surveyed 
region. A similar result has been suggested for the Coma cluster by Andreon \& 
Pell\'o (2000) from the comparison of heterogeneous near--infrared LFs computed 
by different authors on different portions of the Coma cluster. In the optical 
$R$ band, Driver, Couch \& Phillipps (1998) and Secker, Harris \& Plummer (1997)
find a  large variation in the ratio of dwarf to giant galaxies as a function of
the  clustercentric distances, with dwarfs more numerous at large ($r>0.56$ Mpc) 
distances (or lower density regions). This trend is in agreement with the
dependency of the slope of the LF with the density in the $\sim65$ clusters  studied
by Garilli, Maccagni \& Andreon (1999).  Due to the small {\it Hubble Space
Telescope} field of view,  it is not possible to measure whether the found
variation of the LF is due to a change in the relative number  density of galaxies
of each morphological type (i.e. LFs of the morphological  types are universal) or
rather to luminosity changes of the individual
galaxies (in that  case the LFs of each
morphological type depend on location), and should be  deferred until {\it Hubble
Space Telescope} images for a larger field will be  hopefully taken. In the
optical, the bright part of the luminosity functions of the morphological type are
found not to depend on the considered environments and  differences in the optical
LF are found mainly due to differences in the cluster  morphological composition
(Binggeli et al. 1986, Jerjen \& Tammann 1997;  Andreon 1998).

The spatial dependence of the AC\,118 near--infrared LF implies a luminosity 
segregation: bright galaxies are found preferentially in the cluster center. 
There are several claims (e.g. Driver, Couch \& Phillipps 1998; Secker, Harris \&
Plummer 1997) of a luminosity segregation for a few rich clusters, but this evidence is 
restricted to optical bands. In this paper, this evidence is extended to a near—
infrared band. Since the near--infrared luminosity traces stellar mass (Bruzual 
\& Charlot 1993), this result implies a mass segregation more tightly than under 
the usual assumption than optical luminosity traces mass: here we show directly 
that massive galaxies are found preferentially in the cluster center. Thus, more 
massive clumps are more tightly bound to the clusters, which is a general 
outcome of the simulations of a hierarchical universe (Kauffmann et al. 1997). 
The hostile cluster environment plays a role in shaping the AC\,118 LF but only 
at small clustercentric radii (or high density), since the outer region AC\,118 
LF is quite similar to the field FL.

A possible not--ubiquitous cluster LF, i.e. varying with cluster radius,
implies a dependency of the LF parameters
($M^*, \alpha$) on the surveyed area and suggests caution in performing
cosmological tests involving $M^*$ as a standard candle, or in studying the
galaxy evolution through a change in the best fit LF parameters. Recently, De
Propris et al. (1999) report a brightening of $M^*$ with increasing redshift
for their cluster sample, which is characterized by a surveyed area (in
Mpc$^2$) varying with redshift in a complex way due to the variety of field of
views of used instruments and sampled redshifts. They average a few cluster 
at each redshift, in order to reduce the impact of a not--ubiquitous cluster LF.
Typically, areas range from 1.5 to 3.5 Mpc. However, 
the problem of a not-ubiquitous LF cannot probably be
circumvented by choicing a fixed area in the cluster rest--frame because 
clusters have not standard size and shapes. Given the importance of determination
of the evolution of $M^*$, we consider useful to check on a large sample of
clusters the claim of a possible not-ubiquitous cluster near--infrared LF.


\begin{acknowledgements}
This work is part of a collaboration with M. Arnaboldi, G. Busarello, M.
Capaccioli, G. Longo, P. Merluzzi and G. Theureau. E. Bertin and N. Devillard
are warmly thanked for their user friendly software (SEx and Eclipse) and their
exhaustive answers to my detailed questions on technicalities about 
their software. Y. Mellier and E. Bertin helped
me with setting SEx for the AC\,118 image. I thank D. Maccagni for a careful
reading of this manuscript and useful suggestions. A special thank goes to the 
anonymous referee for his very careful work and interesting feedback. 
The near--infrared observations presented in this paper have been taken during the NTT
guaranteed time of Osservatorio di Capodimonte. This work is dedicated to the
memory of my grand--parents, Antonio e Maria Rapposelli.
\end{acknowledgements}

\newpage

\begin{deluxetable}{lcccc}
\tablecolumns{4}
\tablewidth{0pc}
\small
\tablecaption{The data}
\tablehead{\colhead{} & \colhead{AC\,118} & \colhead{HDF-S-1} & \colhead{HDF-S-
2} \\}
\startdata
Exposure time (min)	& 265	& 180	& 300	\nl
Seeing	(FWHM, arcsec)	& 0.75	& 0.90	& 0.96	\nl
Fully corrected noise$^b$ (mag arcsec$^{-2})$ & 24.0 & 24.0 & 24.2 \nl
Complet. mag ($\phi=4.^"4$)	& 20.5 & 20.5	& 21.0 \nl
Galact. lat. (degree)	& -81	& -49	& -49	\nl
Galact. abs. $E(B-V)^a$ & 0.013 & 0.028 & 0.028 \nl
\enddata
\tablecomments{The full SOFI field of view is 
$5\times5$ arcmin. The used field of view is 4'52"$\times$4'52. \\
\null$^a$ Color excess has been measured using COBE/DIRBE maps 
(Schlegel, Finkbeiner \& Davis 1998).
Galactic absorption in $K_s$ is $A_K=0.2 E(B-V)$ (Allen, 1955) \\
\null$^b$ The sky noise is measured as the dispersion of
adjacent pixels in the background, once the image is binned 
in pixels of 1 arcsec for reducing the correlation between
adjacent pixels in the HDF-S images.}
\end{deluxetable}

\begin{deluxetable}{lcccc}
\tablecolumns{4}
\tablewidth{0pc}
\small
\tablecaption{Best Fit values}
\tablehead{\colhead{Region} & \colhead{$K^*$} & \colhead{$\alpha$} & 
\colhead{$\phi$} & 
\colhead{$\chi^2$}\\}
\startdata
Global		& 15.32	& -1.18	& 31.21	& 7.21 	\nl
Main clump	& 16.59	& -0.47	& 30.74 & 4.33	\nl
SW clump	& 17.02	& -0.91	& 19.24 & 10.05	\nl
outer region	& unconstrained	& -1.69	& 0.21	& 10.01	\nl
\enddata
\tablecomments{All fits have 9 degrees of freedom and three free 
parameters, of which two ($K^*$ and $\alpha$) are interesting.}
\end{deluxetable}

\newpage

\begin{figure}
\vbox{
\epsfxsize=15cm
\epsfbox[35 160 560 683]{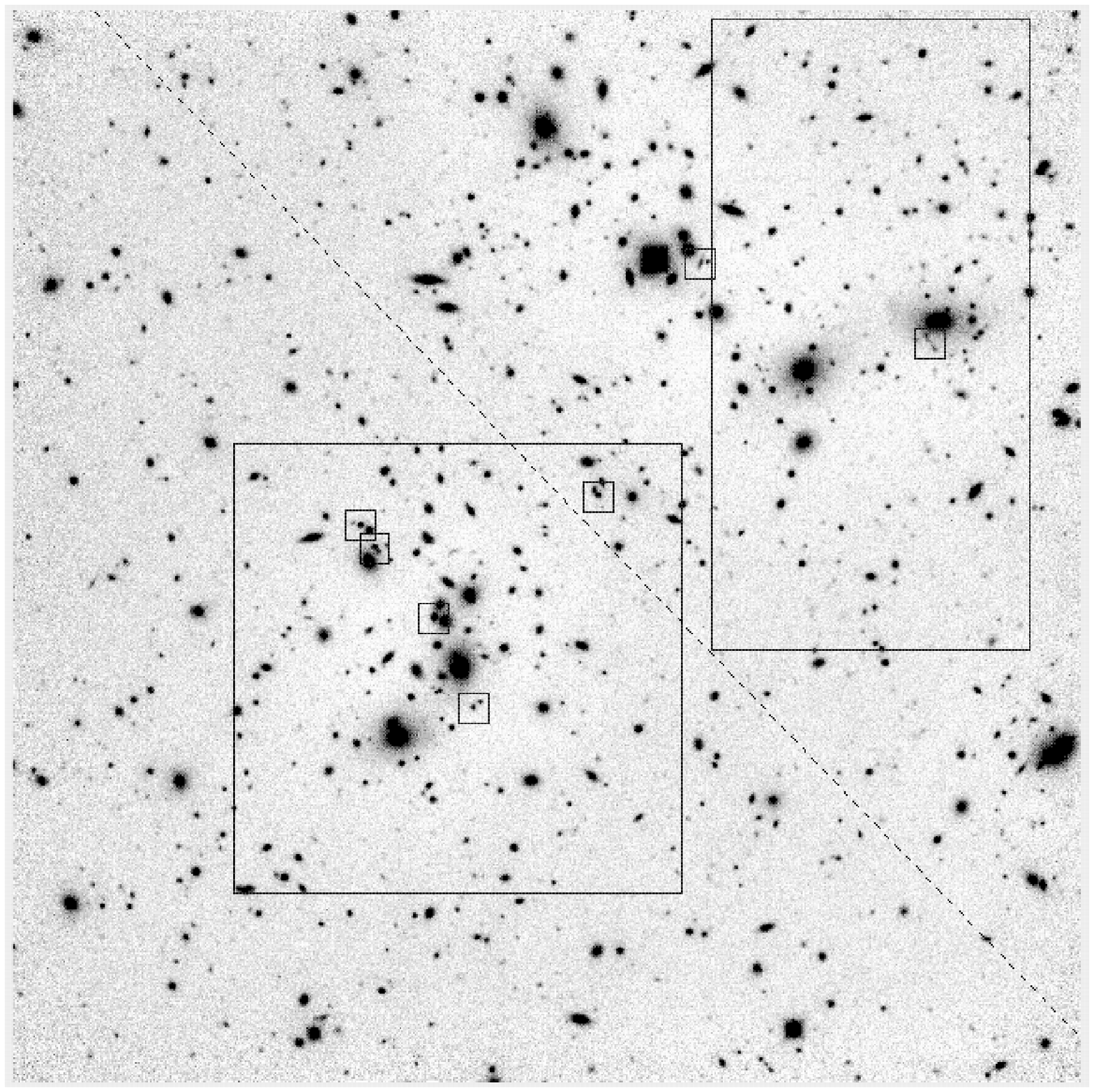}%
\hfill\null}
\vbox{
\epsfxsize=15cm
\epsfbox[35 358 577 434]{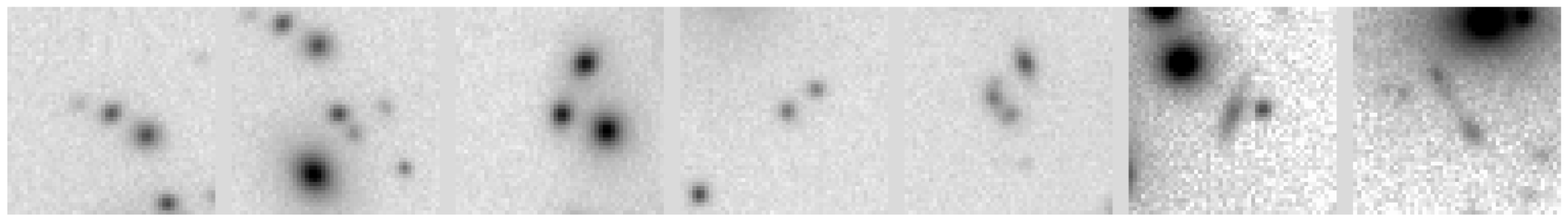}%
}
\caption[h]{The $K_s$ band image of AC\,118. The field of view is
$\sim5\times5$ arcmin. North is up and East is to the left. The large boxes
delimit the main and North--West regions (large square and rectangle,
respectively). The ``outer region" (see text) is the area below the slanted line
and outside the large square.  Small boxes mark the objects blended in Barger 
et al. (1996) image (the 6 small boxes at East) or gravitational lenses (the two small boxes 
at West). All them are magnified in the lower panel. 
The field of view of each zoom is 15
arcsec on side.}
\end{figure}

\begin{figure}
\hbox{
\epsfxsize=8cm
\epsfbox[30 170 570 700]{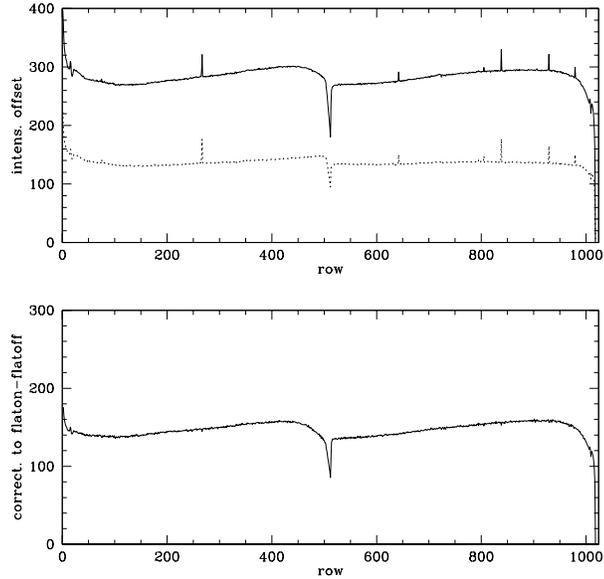}%
\hfill\null}
\caption[h]{{\it Upper panel:} zero--level of flaton (solid
line) and flatoff (dotted line). Spikes are due to bad pixels,
and do not reflect real variations of the zero--level.
{\it Lower panel:} difference of the two zero--levels.
}
\end{figure}

\begin{figure}
\hbox{
\epsfxsize=8cm
\epsfbox[40 190 470 590]{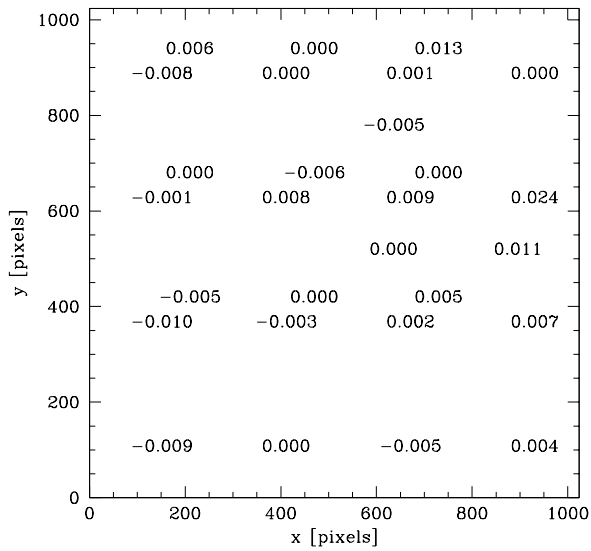}%
\hfill\null}
\caption[h]{
Illumination correction. Values indicate
the deviation from a perfect flat-fielding (i.e. from 1 everywhere) and 
are located where measurements were performed. Typical measurement
errors are 0.005}
\end{figure}

\begin{figure}
\hbox{
\epsfxsize=8cm
\epsfbox[20 170 570 695]{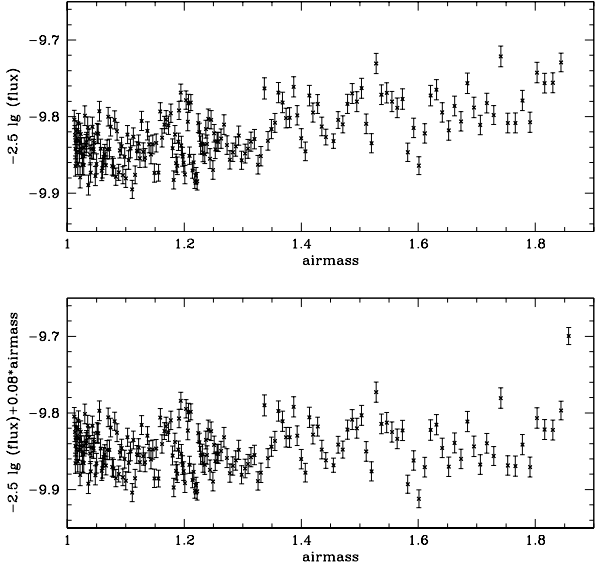}%
\hfill\null}
\caption[h]{{\it Upper panel:}
Airmass dependence measured by a possible elliptical galaxy
of AC\,118. {\it Lower panel:} instrumental magnitude, airmass corrected,
of the same galaxy. Notice the small scatter (0.03 mag) almost
entirely due to photometric errors on the single measure.}
\end{figure}


\begin{figure}
\hbox{
\epsfxsize=8cm
\epsfbox[30 190 470 590]{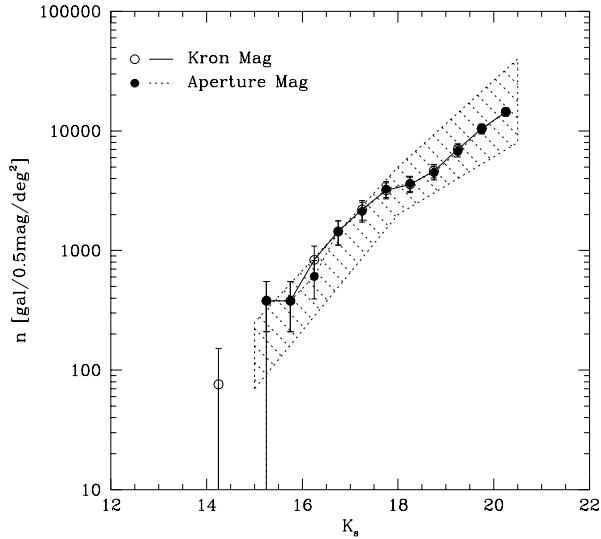}%
\hfill\null}
\caption[h]{Galaxy counts adopting Kron (open points) and aperture
(filled points) magnitudes in the HDF-S line of sight.
Counts are identical almost at all magnitudes, and in particular at
the bridge magnitude $K_s=18$ mag. The hashed
region delimits the locus occuped by literature galaxy counts in the $K$ band, 
as estimated from the compilation presented in Figure 1 of McCracken et al. (2000).
The large amplitude of this region is due in part to heterogeneity of the 
data and reduction of the compared works.
}
\end{figure}


\begin{figure}
\hbox{
\epsfxsize=8cm
\epsfbox[60 195 570 695]{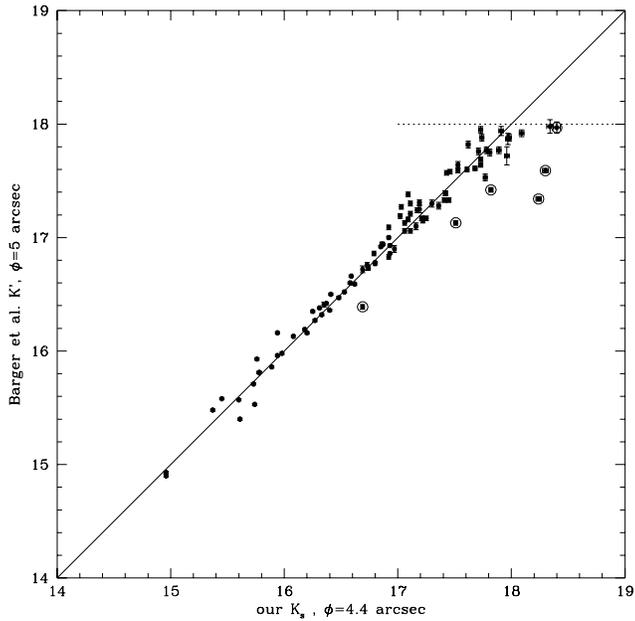}%
\hfill\null}
\caption[h]{Comparison between Barger et al. (1996) $K'$ magnitude
in a 5 arcsec aperture with our $K_s$ magnitudes in a 4.4 arcsec 
aperture for common objects. The line is the ``one--to--one" relation,
not a fit to the data. Outliers to the right
of the line (marked with a circle) are pairs of objects,
blended in Barger et al. (1996) image but resolved as separated
components in our image. The horizontal dotted line marks
the Barger et al. (1996) catalog limit.}
\end{figure}

\begin{figure}
\hbox{
\epsfxsize=8cm
\epsfbox[35 190 465 590]{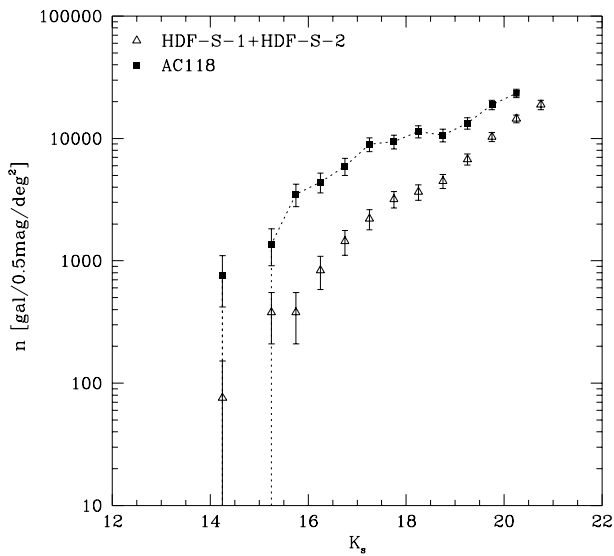}%
\hfill\null}
\caption[h]{Raw galaxy counts in the HDF-S fields and
toward AC\,118. Errorbars are simply taken as $\sqrt n$. }
\end{figure}

\newpage

\begin{figure}
\hbox{
\epsfxsize=8cm
\epsfbox[50 190 465 625]{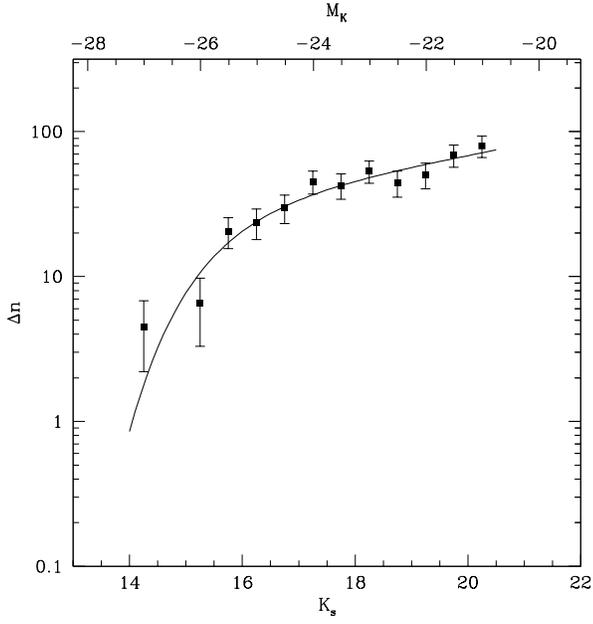}%
\hfill\null}
\caption[h]{AC\,118 global luminosity function. Errorbars
are $\pm1\sigma$ and take into account both
Poissonian and non--Poissonian fluctuations, i.e. include the cosmic
variance of background galaxy counts. Both apparent and
absolute magnitude scales are presented, as in most of the
following figures.}
\end{figure}

\begin{figure}
\hbox{
\epsfxsize=8cm
\epsfbox[60 200 515 590]{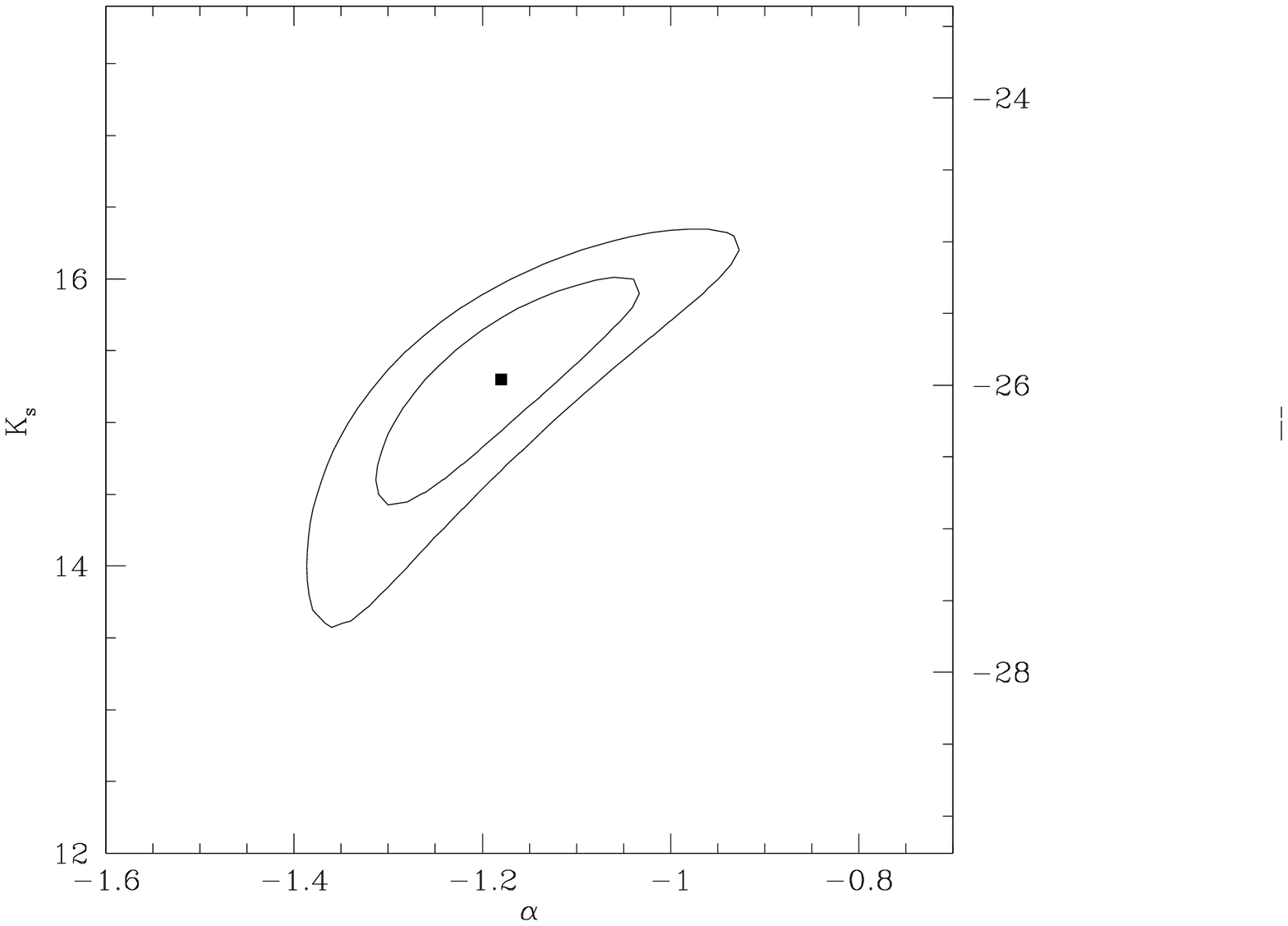}%
\hfill\null}
\caption[h]{68 \% and 95 \% confidence levels
of the best fitting Schechter parameters for the global 
AC\,118 LF. The filled square marks the best fit.}
\end{figure}

\begin{figure}
\hbox{
\epsfxsize=8cm
\epsfbox[25 175 465 595]{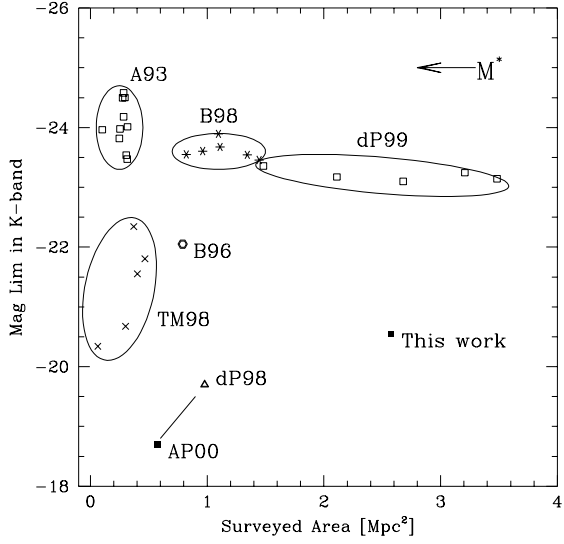}%
\hfill\null}
\caption[h]{Rest--frame surveyed area vs. absolute magnitude limits for previous
determination of the cluster near--infrared
LF. LFs presented in Aragon--Salamanca et al. (1993),
Barger et al. (1998), Barger et al. (1996), de Propris et al. (1998; 1999),
Trentham \& Mobasher (1998), Andreon \& Pell\'o (2000)
are denoted as AE93, B98, B96, dP98, dP99, TM98 and AP00, respectively. The 
characteristic
magnitude, $M^*$, of the Schechter LF is also marked. The line connect two
different LF determinations of the Coma LF.}
\end{figure}

\newpage

\begin{figure}
\hbox{
\epsfxsize=8cm
\epsfbox[50 190 465 625]{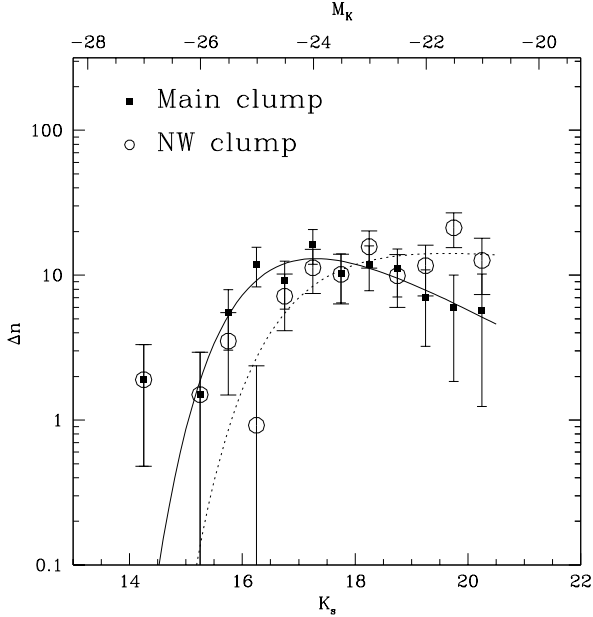}%
\hfill\null}
\caption[h]{AC\,118 LF of the two clumps. Filled squares (open circles) mark
the LF computed for the main (NW) clump. Errorbars are
as in Figure 8. The two curves are the best Schechter fits (solid line
for the main clump, dotted line for the NW clump).
}
\end{figure}

\begin{figure}
\hbox{
\epsfxsize=8cm
\epsfbox[60 200 515 590]{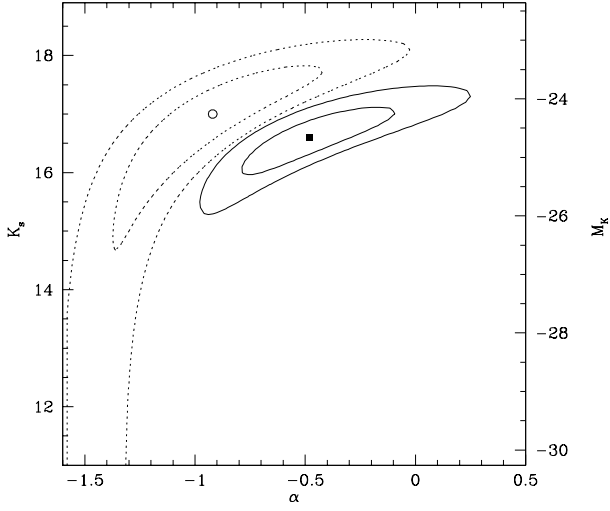}%
\hfill\null}
\caption[h]{68 \% and 95 \% confidence levels
of the best fitting Schechter parameters
for the main clump (solid lines) and NW clump (dotted lines) of AC\,118.
The filled square and the open circle marks the best fit values.}
\end{figure}

\begin{figure}
\hbox{
\epsfxsize=8cm
\epsfbox[50 190 465 625]{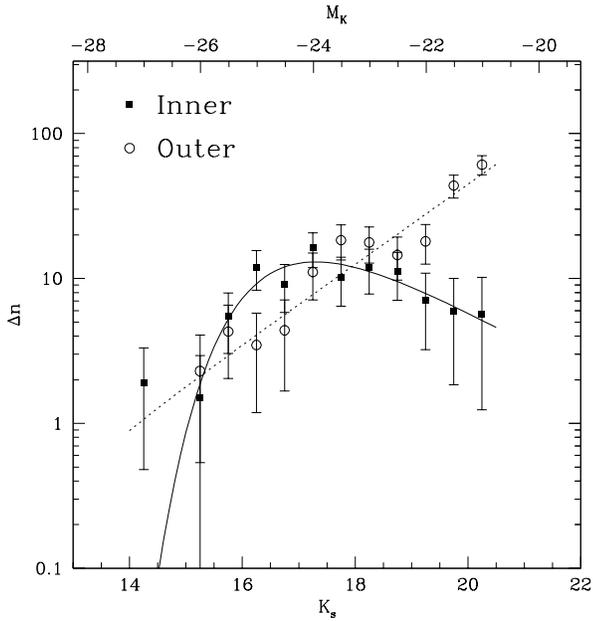}%
\hfill\null}
\caption[h]{AC\,118 LF of the inner and outer region. Solid squares (open 
circles) mark
the LF computed for the inner (outer) region. Errorbars are
as in Figure 8. The two curves are the best Schechter fits (solid line
for the inner region, dotted line for the outer region).
}
\end{figure}

\begin{figure}
\hbox{
\epsfxsize=8cm
\epsfbox[60 200 515 590]{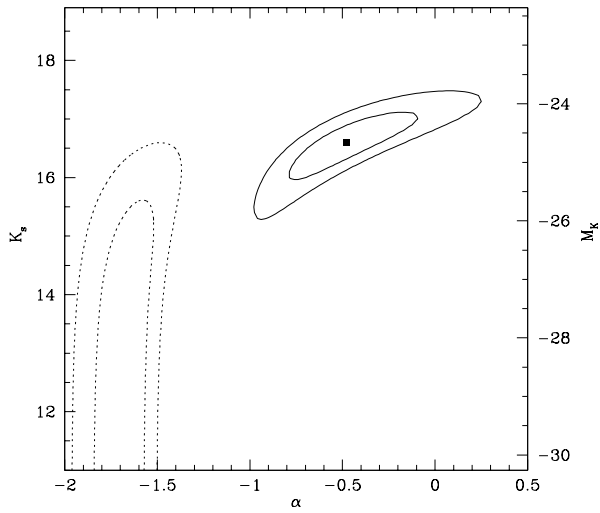}%
\hfill\null}
\caption[h]{68 \% and 95 \% confidence levels
of the best fitting Schechter parameters
for the inner region (solid lines) and outer region (dotted lines) of AC\,118. 
The filled square marks the best fit of the inner region LF. The best fit for
the outer region is undetermined.}
\end{figure}

\begin{figure}
\hbox{
\epsfxsize=8cm
\epsfbox[60 200 515 590]{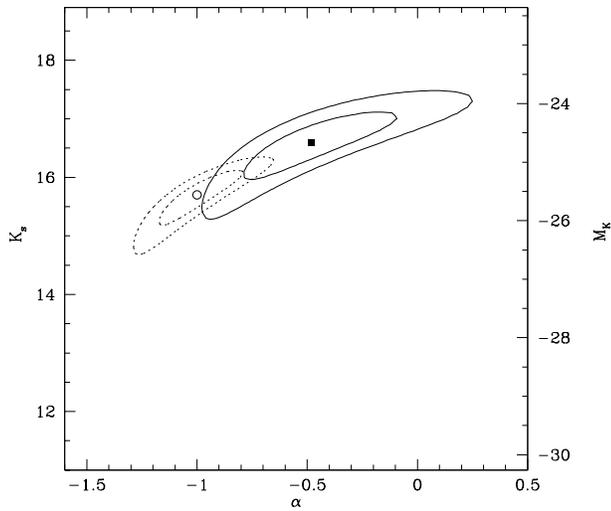}%
\hfill\null}
\caption[h]{68 \% and 95 \% confidence levels
of the best fitting Schechter parameters
for the inner region of AC\,118 (solid lines) and the composite LF of 3 clusters 
at $z\sim0.3$, including AC\,118, presented in Barger et al. (1996) (dotted 
lines). 
The filled square and the open circle mark the best fit values.}
\end{figure}

\begin{figure}
\hbox{
\epsfxsize=8cm
\epsfbox[50 190 465 590]{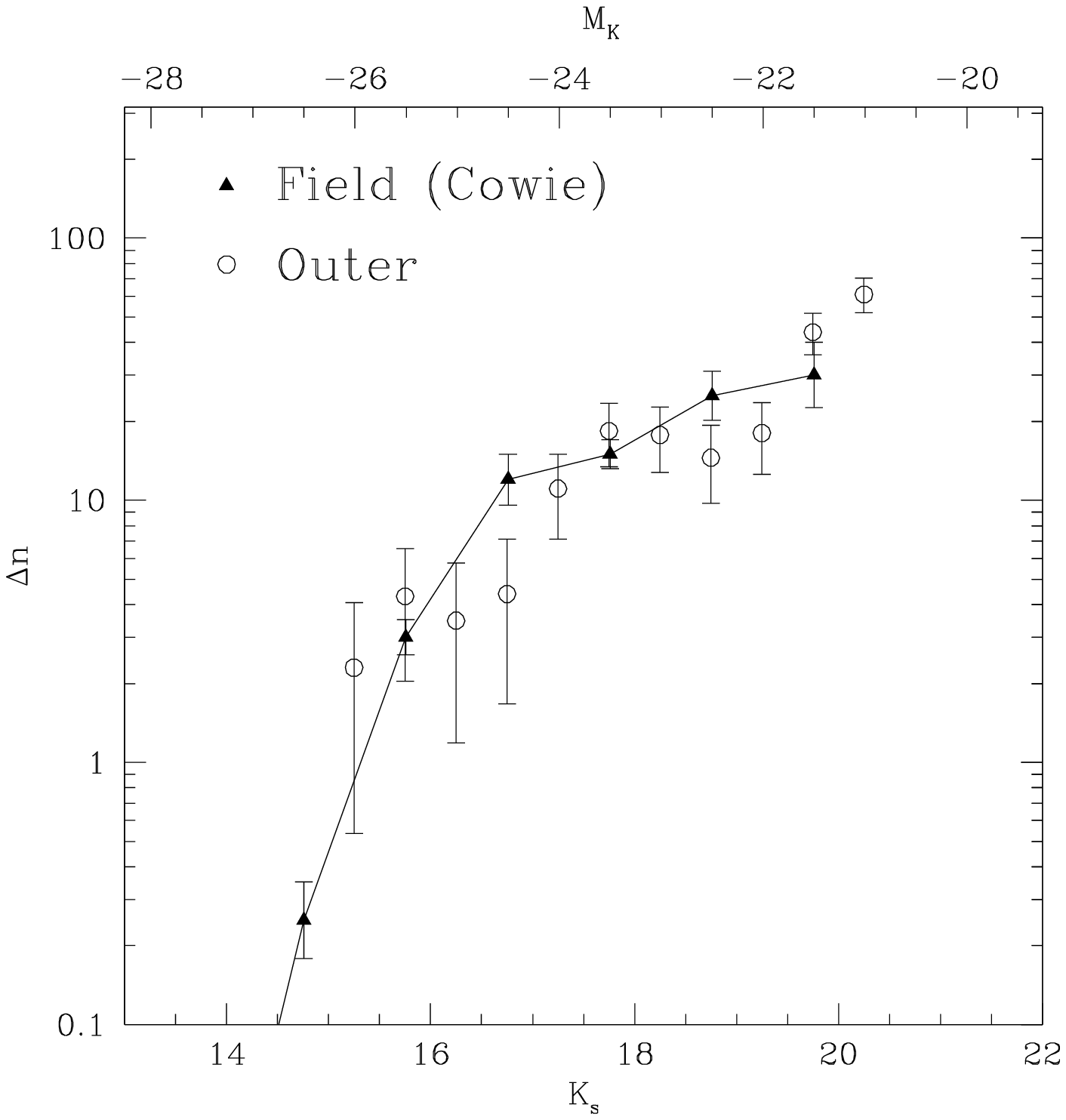}%
\hfill\null}
\caption[h]{
LF in the outer region of AC\,118 (open points) and in the field at $0.2<z<0.6$
(from Cowie et al. 1996, filled triangles connected by a solid line). 
The field LF has been vertically shifted to match the AC\,118 LF. 
}
\end{figure}


\newpage

\begin{thebibliography}{}

\bibitem[Abell 1958]{1958ApJS....3..211A} 
Abell G. O., 1958, ApJ 3, 211 

\bibitem[]{}
Allen C. W., 1955, Astrophysical quantities,
University of London, Athlone Press, 1st edition 

\bibitem[]{}
Andreon S., 1993, in 5$^{th}$ ESO/ST-ECF Data Analysis
Workshop, eds P.J. Grosbol and R.C.E. Ruijsscher, Garching bei M\"unchen,
p. 219

\bibitem[]{}
Andreon S., 1998, A\&A 336, 98

\bibitem[]{}
Andreon S., Cuillandres J.-C., 2000, in preparation


\bibitem[]{}
Andreon S., Pell\'o R., Davoust E., Dom\' \i nguez R., Poulain P., 2000, A\&AS, 
141, 113 


\bibitem[]{}
Andreon S. \& Pell\'o R., 2000, A\&A, 353, 479

\bibitem[Aragon-Salamanca Ellis Couch \& Carter 1993]{1993MNRAS.262..764A} 
Aragon-Salamanca A. , Ellis R. S., Couch W. J. \& Carter D., 1993, MNRAS 262, 
764 

\bibitem[Avni 1976]{1976ApJ...210..642A} 
Avni Y., 1976, ApJ 210, 642 

\bibitem[Balogh et al. 1997]{1997ApJ...488L..75B} 
Balogh M. L., Morris S. L., Yee H. K. C., Carlberg R. G. \& Ellingson E. 1997, 
ApJ 488, L75 

\bibitem[Balogh et al. 1998]{1998ApJ...504L..75B} 
Balogh M. L., Schade D. , Morris S. L., Yee H. K. C., Carlberg R. G. \& 
Ellingson E. 1998, 
ApJ 504, L75 

\bibitem[Barger et al. 1996]{1996MNRAS.279....1B} 
Barger A. J., Aragon-Salamanca A., Ellis R. S., Couch W. J., Smail I., Sharples 
R. 
M., 1996, MNRAS 279, 1

\bibitem[Barger et al. 1998]{1998ApJ...501..522B}
Barger A. J., et al. 1998, ApJ 501, 522 
 
\bibitem[]{} 
Bertin E., \& Arnouts S. 1996, A\&AS, 117, 393

\bibitem[]{}
Binggeli B., 1986, in {\it Nearly Normal Galaxies}, ed. S. Faber,
(Springer Verlag), p. 195

\bibitem[Binggeli Sandage \& Tammann 1985]{1985AJ.....90.1681B} 
Binggeli, B., Sandage, A. \& Tammann, G. A. 1985, AJ 90, 1681 

\bibitem[]{}
Binggeli B., Sandage A., Tammann G. A. 1988, Ann. Rev. Astron. Astrophys. 26, 
509

\bibitem[Butcher \& Oemler 1978]{} 
Butcher H. \& Oemler A. Jr., 1978, ApJ 219, 18 


\bibitem[Butcher \& Oemler 1984]{1984ApJ...285..426B} 
Butcher H. \& Oemler A. Jr., 1984, ApJ 285, 426 

\bibitem[Bruzual A. \& Charlot 1993]{1993ApJ...405..538B} 
Bruzual A. G., Charlot S. 1993, ApJ 405, 538 

\bibitem[Colina Holfeltz \& Ritchie 1998]{1998nvlt.proc...36C} Colina L., 
Holfeltz S. \& Ritchie C. 1998, in NICMOS and the VLT, 
ESO Conference and Workshop Proceedings 55, Wolfram 
Freudling and Richard Hook eds., p. 36 

\bibitem[Couch \& Newell 1984]{1984ApJS...56..143C}
Couch W. J. \& Newell E. B. 1984, ApJS 56, 143 

\bibitem[Couch \& Sharples 1987]{1987MNRAS.229..423C} 
Couch W. J. \& Sharples R. M. 1987, MNRAS 229, 423 

\bibitem[Couch et al. 1998]{1998ApJ...497..188C} 
Couch W. J., Barger A. J., Smail I. , Ellis R. S., Sharples R. M. 1998, ApJ 497, 
188 

\bibitem[Cowie Songaila Hu \& Cohen 1996]{1996AJ....112..839C} 
Cowie L. L., Songaila A. , Hu E. M. \& Cohen J. G. 1996, AJ 112, 839 

\bibitem[]{}
Da Costa L., Nonino M., Rengelink R., Zaggia S., Benoist C. et al.,
2000, A\&A, submitted (astro-ph/9812105)

\bibitem[De Propris Eisenhardt Stanford \& Dickinson 1998]{1998ApJ...503L..45D} 
De Propris R., Eisenhardt P. R., Stanford S. A., Dickinson M., 1998, ApJ 503, 
L45 

\bibitem[De Propris et al. 1999]{1999AJ....118..719D}
De Propris R., Stanford S. A., Eisenhardt P. R., Dickinson M., Elston R.,
1999, AJ, 118, 719 

\bibitem[]{}
Devillard N., 1997, The Messenger 87, 19

\bibitem[Driver \& Phillipps 1996]{1996ApJ...469..529D} 
Driver S. P., Phillipps S., 1996, ApJ 469, 529 

\bibitem[Driver Couch \& Phillipps 1998]{1998MNRAS.301..369D}
Driver S. P., Couch W. J., Phillipps S 1998, MNRAS 301, 369 

\bibitem[Ettori \& Rosat 1999]{1999MNRAS.305..834E} 
Ettori S. \& Rosat A. C. F. 1999, MNRAS 305, 834 

\bibitem[Gaidos 1997]{1997AJ....113..117G}
Gaidos E. J. 1997, AJ 113, 117

\bibitem[]{}
Garilli B., Maccagni D., Andreon S., 1999, A\&A, 342, 408

\bibitem[]{}
Gehrels N., 1986, ApJ 303, 336

\bibitem[]{}
Jerjen H., Tammann G., 1997, A\&A 321, 713

\bibitem[Huang et al. 1997]{1997ApJ...476...12H} 
Huang J.-S., Cowie L. L., Gardner J. P., Hu E. M., 
Songaila A. \& Wainscoat R. J., 1997, ApJ 476, 12 

\bibitem[Kauffmann Nusser \& Steinmetz 1997]{1997MNRAS.286..795K} 
Kauffmann G., Nusser A., Steinmetz M. 1997, MNRAS 286, 795 

\bibitem[]{}
Koo D. C., Kron R.G., 1992, ARA\&A 30, 613

\bibitem[]{}
Kron R. G. 1980, ApJS 43, 305 

\bibitem[Lumsden et al. 1997]{1997MNRAS.290..119L} 
Lumsden S. L., Collins C. A., Nichol R. C., Eke V. R., Guzzo L. 1997, MNRAS 290, 
119 

\bibitem[Mackay et al. 1998]{1998SPIE.3354...14M}
Mackay C. D., Beckett M., Mcmahon R., Parry I., Piche F.,
Ennico K., Kenworthy M., Ellis R., Aragon-Salamanca A.,
Proc. of SPIE, 3354, 14 

\bibitem[McCracken et al. 2000]{2000MNRAS.311..707M} 
McCracken, H. J., Metcalfe, N., Shanks, T., Campos, A., Gardner, J. P. \& Fong, R. 2000, 
\mnras, 311, 707 

\bibitem[]{}
Moorwood A., Cuby J.-G., Lidman C., 1998, The Messenger, 91, 9

\bibitem[Oemler 1974]{1974ApJ...194....1O} 
Oemler, A. , Jr. 1974, \apj, 194, 1 


\bibitem[]{}
Ostriker G., 1993, ARA\&A 31, 589

\bibitem[Persson et al. 1998]{1998AJ....116.2475P} 
Persson S. E., Murphy D. C., Krzeminski W., Roth M., Rieke M. J. 1998, AJ 116, 
2475 

\bibitem[Phillipps Driver Couch \& Smith 1998]{1998ApJ...498L.119P} 
Phillipps S., Driver S. P., Couch W. J., Smith, R. M. 1998, ApJ 498, L119 

\bibitem[Press et al. 1993]{Press93} 
Press W. H., Teukolsky S. A., Vetterling W. T., 
Flannery B. P., 1993, Numerical Receipes, Cambridge
University Press, Cambridge

\bibitem[Schechter 1976]{1976ApJ...203..297S} 
Schechter P., 1976, ApJ 203, 297 

\bibitem[Schlegel Finkbeiner \& Davis 1998]{1998ApJ...500..525S} 
Schlegel D. J., Finkbeiner D. P., Davis M. 1998, ApJ 500, 525 

\bibitem[Secker Harris \& Plummer 1997]{1997PASP..109.1377S} 
Secker, J., Harris, W. E. \& Plummer, J. D. 1997, \pasp, 109, 1377 

\bibitem[Simard et al. 1999]{1999ApJ...519..563S} 
Simard L., Koo D., Faber S., Sarajedini V., Vogt N., et al. 1999, ApJ 519, 563 

\bibitem[Smail et al. 1991]{1991MNRAS.252...19S} 
Smail I., Ellis R. S., Fitchett M. J., Norgaard-Nielsen H. U., Hansen L., 
Jorgensen H. E. 
1991, MNRAS 252, 19 

\bibitem[Smail et al. 1997]{1997ApJ...479...70S} 
Smail I., Ellis R. S., Dressler A. , Couch W. J., Oemler A. Jr., Sharples R. M., 
Butcher 
H. 1997, ApJ 479, 70 

\bibitem[Trentham \& Mobasher 1998]{1998MNRAS.299..488T} 
Trentham N., Mobasher B., 1998, MNRAS 299, 488 

\bibitem[Valotto Nicotra Muriel \& Lambas 1997]{1997ApJ...479...90V} 
Valotto C. A., Nicotra M. A., Muriel H., Lambas D. G. 1997, ApJ 479, 90 

\bibitem[Wilson Smail Ellis \& Couch 1997]{1997MNRAS.284..915W} 
Wilson G., Smail I., Ellis R. S., Couch W. J. 1997, MNRAS 284, 915 

\bibitem[Zwicky 1957]{1957moas.book.....Z} Zwicky, F.  1957, 
Morphological astronomy, Berlin: Springer

\end{thebibliography}
\end{document}